\documentclass[iop,apj]{emulateapj}
\usepackage{apjfonts}
\usepackage{graphicx}
\usepackage{hyperref}

\begin{document}

\title{A Universal Transition in Atmospheric Diffusion for Hot Subdwarfs Near
  18,000~K\altaffilmark{1}}
\shorttitle{A Universal Transition Near 18,000~K}

\author{
T.M. Brown\altaffilmark{2}, 
J.M. Taylor\altaffilmark{2},
S. Cassisi\altaffilmark{3},
A.V. Sweigart\altaffilmark{4},
A. Bellini\altaffilmark{2}, 
L.R. Bedin\altaffilmark{5},\\
M. Salaris\altaffilmark{6},
A. Renzini\altaffilmark{5}, and
E. Dalessandro\altaffilmark{7},
}

\altaffiltext{1}{Based on observations made with the NASA/ESA {\it Hubble
Space Telescope}, obtained at the Space Telescope Science Institute, which
is operated by the Association of Universities for Research in Astronomy, 
Inc., under NASA contract NAS 5-26555.  These observations are associated
with program GO-14759.}

\altaffiltext{2}{
Space Telescope Science Institute, 3700 San Martin Drive,
Baltimore, MD 21218, USA;  tbrown@stsci.edu; bellini@stsci.edu; jotaylor@stsci.edu}

\altaffiltext{3}{
INAF -- Osservatorio Astronomico d'Abruzzo, Via Mentore
Maggini s.n.c., I-64100 Teramo, Italy; cassisi@oa-teramo.inaf.it;}

\altaffiltext{4}{
NASA Goddard Space Flight Center, Greenbelt, MD 20771, USA; allen.sweigart@gmail.com}

\altaffiltext{5}{
INAF -- Osservatorio Astronomico di Padova, Vicolo dell'Osservatorio 5,
I-35122 Padova, Italy; luigi.bedin@oapd.inaf.it; alvio.renzini@oapd.inaf.it}

\altaffiltext{6}{
Astrophysics Research Institute, Liverpool John Moores University, Liverpool Science
Park, IC2 Building, 146 Brownlow Hill, Liverpool L3 5RF, UK; M.Salaris@ljmu.ac.uk}

\altaffiltext{7}{
INAF -- Osservatorio Astronomico di Bologna, Via Gobetti 93/3, 40129,
Bologna, Italy; emanuele.dalessandro@oabo.inaf.it}

\submitted{Accepted for publication in The Astrophysical Journal}

\begin{abstract}

In the color-magnitude diagrams (CMDs) of globular clusters, when the
locus of stars on the horizontal branch (HB) extends to hot
temperatures, discontinuities are observed at colors corresponding to
$\sim$12,000~K and $\sim$18,000~K.  The former is the ``Grundahl
jump'' that is associated with the onset of radiative levitation in
the atmospheres of hot subdwarfs.  The latter is the ``Momany jump''
that has remained unexplained.  Using the Space Telescope Imaging
Spectrograph on the {\it Hubble Space Telescope,} we have obtained
ultraviolet and blue spectroscopy of six hot subdwarfs straddling the
Momany jump in the massive globular cluster $\omega$ Cen.  By
comparison to model atmospheres and synthetic spectra, we find that
the feature is due primarily to a decrease in atmospheric Fe
for stars hotter than the feature, amplified by the temperature
dependence of the Fe absorption at these effective temperatures.

\end{abstract}


\section{Introduction}

The well-behaved luminosity of the horizontal branch (HB) makes it an
important standard candle in old populations (e.g., Carretta et
al.\ 2000).  However, the HB locus exhibits many morphological
peculiarities that have been the subject of study for decades. These
include the ``second parameter'' debate over the factors driving the
HB color distribution (see Catelan 2009 for a review), overluminous HB
stars in He-enhanced populations (e.g., Busso et al.\ 2007; Caloi \&
D'Antona 2007), subluminous HB stars beyond the hot end of the
standard HB sequence (D'Cruz et al.\ 1996, 2000; Brown et al.\ 2001), and
luminosity jumps within the HB distribution (Grundahl et al.\ 1998,
1999; Momany et al.\ 2002, 2004).  Even the existence of the HB itself,
representing a range in envelope mass for an approximately constant core
mass, is a reminder of one of astronomy's mysteries -- the dispersion
in mass loss occurring on the red giant branch.

\begin{figure}[h]
\begin{center}
\includegraphics[width=3.2in]{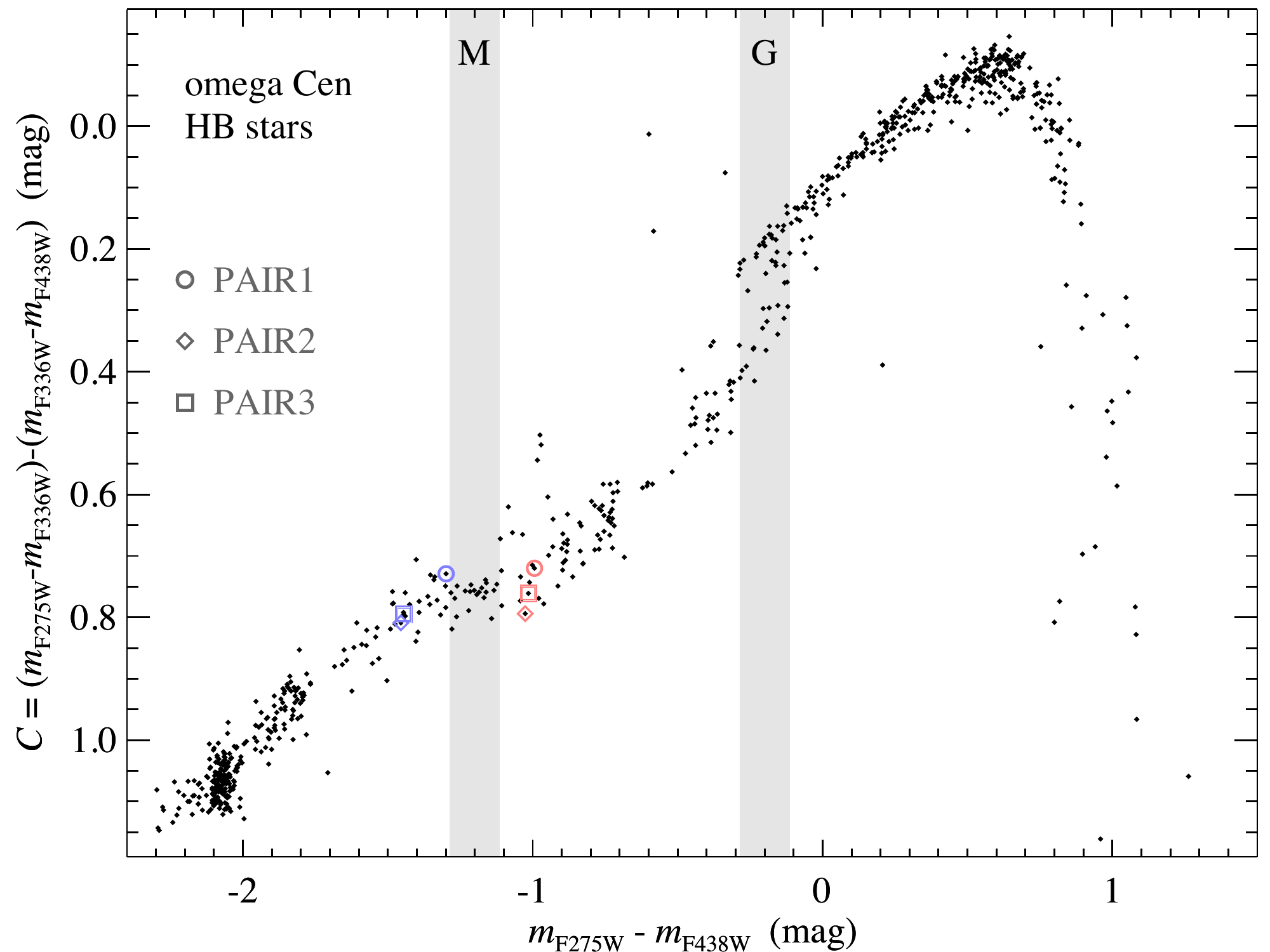}
\end{center}
\caption{The WFC3 photometry ({\it black points}) of HB stars in omega
  Cen (Bellini et al.\ 2017a), shown as a color-color diagram that
  makes it easy to discern the G-jump and M-jump ({\it labeled};
  as in Brown et al.\ 2016).  The ordinate is a color index ($C$)
  that is useful for discerning changes in the stellar
  atmosphere (in the hot stars here but also the cooler stars
  studied in the multiple populations phenomenon).
  For our investigation, we obtained STIS
  near-UV (G230L) and blue (G430L) spectroscopy of six stars ({\it
    colored symbols} matching those in Figure~2)
  that straddle the M-jump, with wavelength coverage
  that includes the three photometric bands employed in this color-color
  diagram (F275W, F336W, and F438W). If the photometry
  here is aligned to a theoretical zero-age HB locus (see Brown et al.\ 2016),
  the gray band
  representing the M-jump spans $\sim$15,000--18,000~K in effective
  temperature.}
\end{figure}

\begin{figure*}[t]
\begin{center}
\includegraphics[width=6.5in]{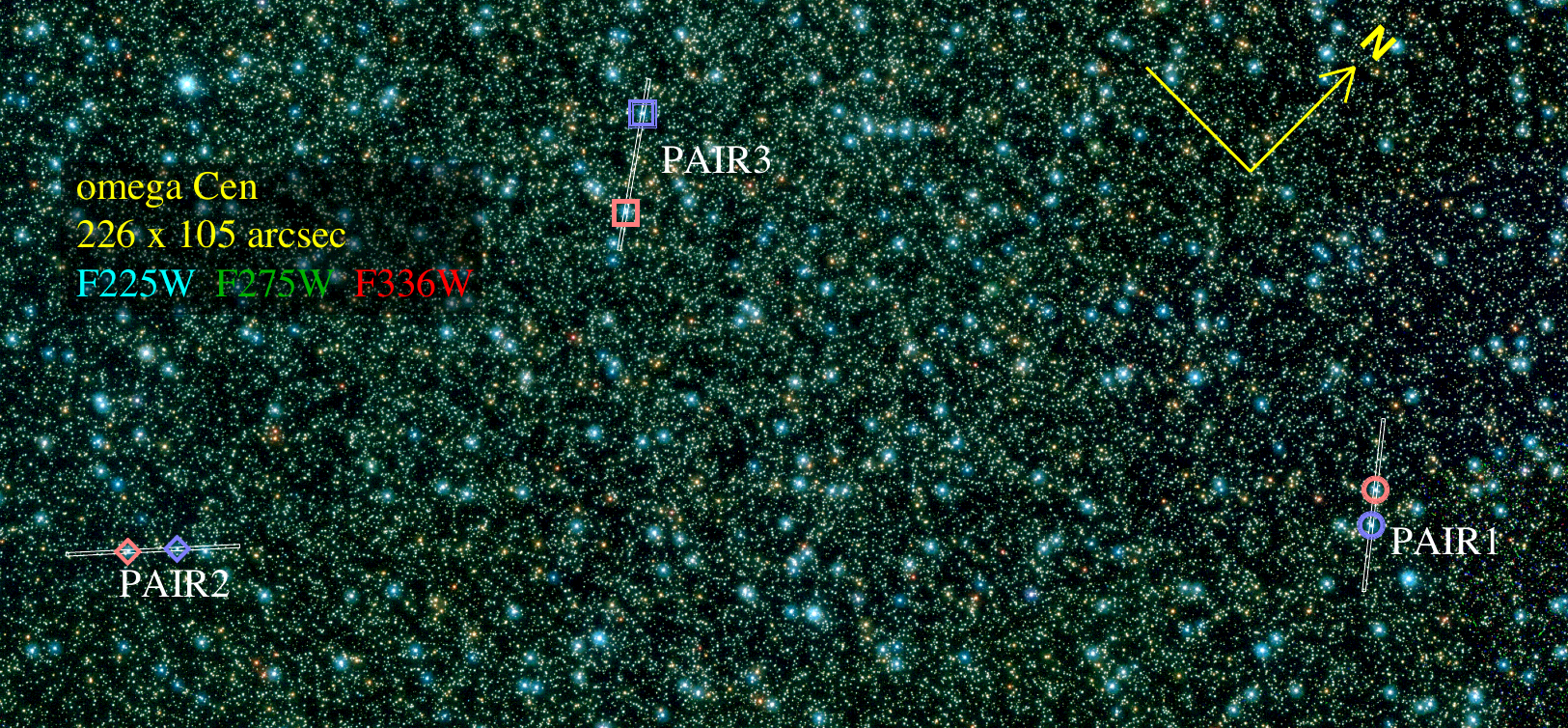}
\end{center}
\caption{ A false-color image of our field, constructed from three
  ultraviolet bands of WFC3 imaging to highlight the hottest stars:
  F225W ({\it blue}), F275W ({\it green}) and F336W ({\it red}).  The
  images come from an {\it HST} calibration program
  (program 11452; PI: Quijano).
  Note that Brown et al.\ (2016) explored HB discontinuities with the
  F275W, F336W, and F438W filters, which motivated
  the wavelength coverage obtained in our spectroscopy here.  A
  $25^{\prime\prime} \times $0\farcs5 box indicates the
  three positions of the STIS slit in our program.  At each slit
  position, a star redward of the M-jump and a star blueward of the
  M-jump were placed within the slit ({\it colored symbols} matching those in
  Figure~1).  In this paper, targets are designated by the slit
  position and this relative color (e.g., Blue\,2 is the blue star in
  the pair at slit position 2).}
\end{figure*}

The {\it Hubble Space Telescope} ({\it HST}) UV Legacy Survey of
Galactic Globular Clusters treasury program (Piotto et al.\ 2015)
enabled a new investigation of these phenomena by providing a large
and homogeneous catalog of UV and blue photometry for over 50 globular
clusters.  The primary goal of the program was an investigation of the
multiple populations phenomenon in globular clusters. The bandpasses
for the program are sensitive to abundance variations in cool stars,
allowing these distinct populations to be cleanly separated and
characterized (e.g., Milone et al.\ 2017a).  A secondary goal of the
program was a characterization of the hot subdwarfs in these clusters.
Brown et al.\ (2016) used this dataset to explore the ``Grundahl
jump'' (hereafter G-jump; Grundahl et al.\ 1998, 1999) and ``Momany
jump'' (hereafter M-jump; Momany et al.\ 2002, 2004) that appear as
discontinuities within the HB distribution near effective temperatures
of $\sim$12,000~K and $\sim$18,000~K (Figure~1).  In the color-color
plane of Figure~1, the stars between the G-jump and M-jump appear to
deviate from the expected location of the theoretical zero-age HB (see
Figures 6 and 7 in Brown et al.\ 2016).  They found that the G-jump
falls at a consistent temperature for 31 of the 33 survey clusters
where it is populated; the only exceptions are NGC~6388 and NGC~6441,
two metal-rich clusters greatly enhanced in helium.  The G-jump is
associated with the onset of radiative levitation for metals and the
gravitational settling of helium in the atmospheres of subdwarfs
hotter than the G-jump (Moehler et al.\ 1999, 2000; Behr 2003; Pace et
al.\ 2006).  Brown et al.\ (2016) explained that the hotter G-jump in
NGC~6388 and NGC~6441 is likely associated with helium enhancement
because it shifts the onset of surface convection to higher
effective temperatures.  They also found that the M-jump is at a
consistent temperature for all 15 of the survey clusters where it is
populated, including the clusters NGC~6388 and NGC~6441.  Given the
variety of cluster parameters in this set (e.g., mass, metallicity,
helium abundance), the consistency of the M-jump demonstrates that it
is clearly a universal atmospheric phenomenon (like the G-jump).
However, with only broad-band photometry, it
was not possible to determine the atmospheric changes associated with
the feature.

To further investigate the nature of the M-jump, we obtained {\it HST}
spectroscopy of hot subdwarfs straddling the feature in the
massive globular cluster $\omega$~Cen (Figure~1).  One may wonder why we chose
such an unusual cluster -- the most massive in the Galaxy, and a striking
example of the multiple populations phenomenon in globular
clusters, with variations in the light elements and iron
(e.g., Bellini et al.\ 2017b; Marino et al.\ 2011).
First, it is worth noting that the hot HB stars are preferentially drawn
from a subset of the $\omega$~Cen populations -- the
bluest of the multiple main sequences (see Cassisi et al.\ 2009).
That said, even with the cluster's distinct sub-populations, the cluster's
composite population is better constrained than
an assorted collection of field subdwarfs.  More importantly,
Brown et al.\ (2016) demonstrated that the G-jump
and M-jump are the products of stellar atmosphere transitions instead
of population distinctions; the transition responsible for the G-jump
can be shifted to hotter effective temperature
at extreme He abundance, but the M-jump is constant even
in the presence of such He enhancements.  For these reasons, the peculiar
populations of $\omega$~Cen are not a concern for this study.
The main motivation for targeting
$\omega$~Cen is that it offers a large sample of relatively nearby
stars straddling the M-jump, with a level of crowding that enables clean
spectroscopy with the Space Telescope Imaging Spectrograph (STIS)
aboard {\it HST}.  In this paper, we analyze the spectra of
six stars (three on each side of the M-jump) to characterize the
atmospheric transitions responsible for the feature.

\begin{figure*}[t]
\begin{center}
\includegraphics[width=6.5in]{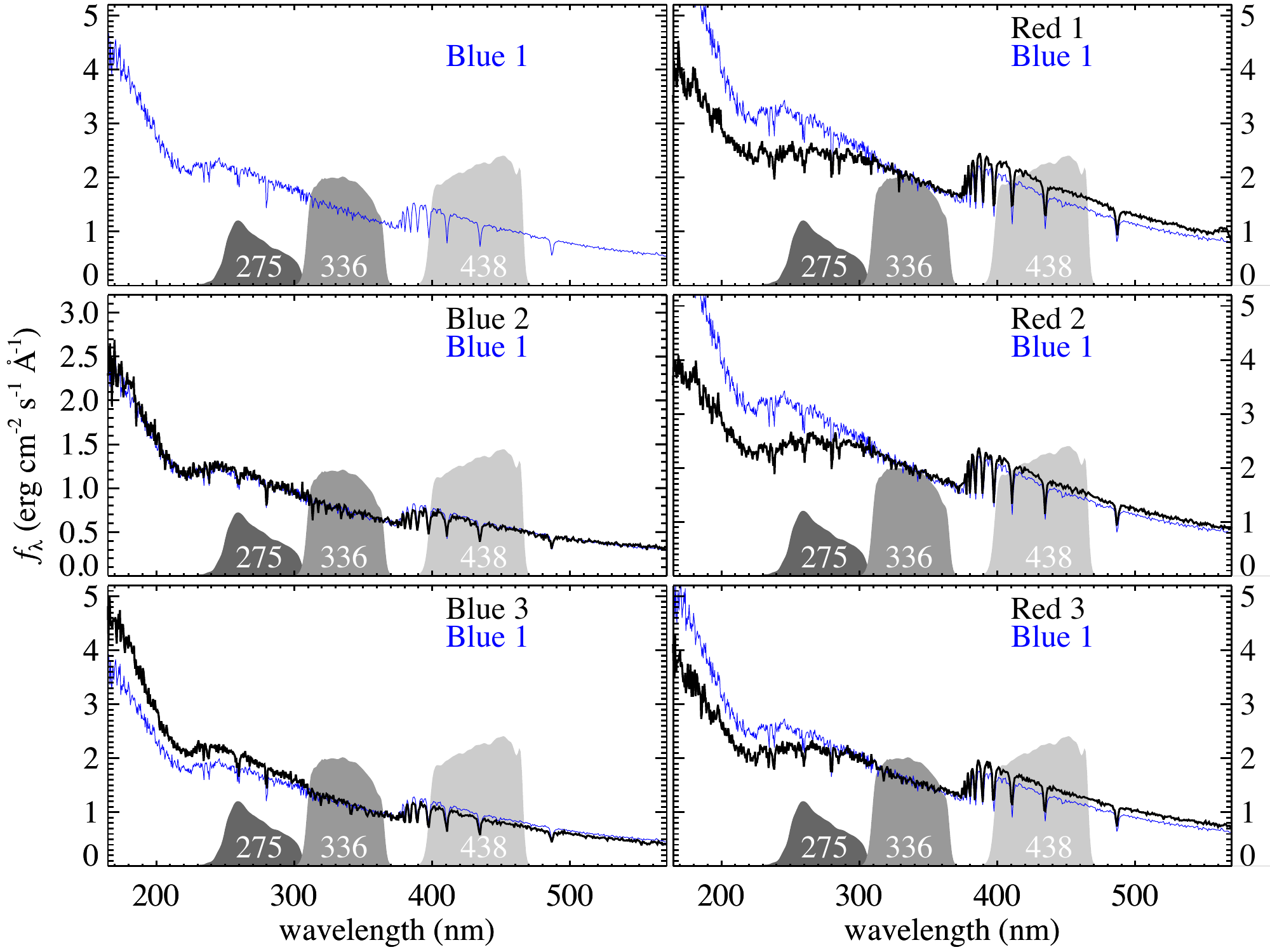}
\end{center}
\caption{The spectra for each of the six stars in our analysis ({\it black
    curves}). The spectrum of Blue\,1 ({\it blue curves}) is overplotted
  in each panel, normalized at 310 -- 360 nm (i.e., F336W) to ease comparisons
  between the spectra.  We also show the bandpasses of the WFC3 filters
  ({\it gray shading}; arbitrary normalization) used in the color-color
  diagram that drove our investigation (Figure~1).
}
\end{figure*}

\begin{table}
  \begin{center}
    \caption{Targets}
    \begin{tabular}{ccccc}
      \tableline
           & R.A.  & Dec. & $m_{\rm F275W} - m_{\rm F438W}$ & $C$ \\
      Name & (J2000) & (J2000) & (mag) & (mag) \\
      \tableline
      Blue\,1 & 13 26 36.45 & -47 28 28.54 & -1.30 & 0.73 \\
      Red\,1  & 13 26 36.76 & -47 28 24.49 & -0.99 & 0.72 \\
      Blue\,2 & 13 26 48.23 & -47 30 32.35 & -1.46 & 0.81 \\
      Red\,2  & 13 26 48.70 & -47 30 37.56 & -1.03 & 0.79 \\
      Blue\,3 & 13 26 47.90 & -47 29 00.64 & -1.45 & 0.80 \\
      Red\,3  & 13 26 47.08 & -47 29 12.47 & -1.02 & 0.76 \\
      \tableline
    \end{tabular}
  \end{center}
\end{table}

\section{Data}

Our program obtained six orbits of {\it HST}/STIS spectroscopy in $\omega$~Cen,
with one orbit of near-UV (G230L) and one orbit of blue (G430L) exposures
at each of three slit positions (Figure~2).  Each position of the
52$^{\prime\prime} \times$0\farcs5 slit was
chosen to sample a pair of stars straddling the M-jump in $\omega$~Cen
(Figure~1 and Table~1);
note that the slit spans 52$^{\prime\prime}$ in the G430L exposures,
but is limited by the detector to 25$^{\prime\prime}$ in the G230L exposures.
Placement of each stellar pair along the slit midline was facilitated by
archival {\it HST} imaging (providing excellent relative astrometry)
and acquisition on a bright isolated star (placing the relative astrometry
in an absolute frame).  For each slit position, three individual exposures
were obtained for each of the two gratings (G230L and G430L), with dithering
along the slit to enable removal of detector artifacts.
A link to the data is provided here:
  \href{http://dx.doi.org/10.17909/T9BT2R}{10.17909/T9BT2R}.
  
The STIS pipeline provides a variety of corrections (see Bostroem \& Proffitt
2011), but the reduction steps for the near-UV and blue spectroscopy are
somewhat distinct, given the different detectors involved.  G430L data are
obtained with the STIS charge coupled device (CCD), which has suffered
considerable radiation damage during its 20 years in flight.  This
damage has reduced the charge transfer efficiency (CTE) and greatly
increased the number of hot pixels.  To account for these issues, we
processed the data with contemporaneous darks and biases, and then
performed a pixel-based correction for CTE losses using the
{\sc stis\_cti} routine, which is based on the algorithms
used on the {\it HST} imaging instruments
(Anderson \& Bedin 2000).  The resulting frames were
shifted to a common position (accounting for the dithers along the
slit) and combined with masking of cosmic rays, hot pixels, and dead
pixels.  The blue spectra were then extracted with the {\sc x1d}
routine in the {\sc stistools} software package. 
G230L data are obtained with a
STIS multi-anode multi-channel array, where CTE is not an issue, there
is little sensitivity to cosmic rays, and the evolution in defective
pixels is much slower over time.  The data were dark and bias
corrected, shifted to a common frame, combined with masking of
defective pixels, and extracted with the {\sc x1d} routine. The
near-UV and blue spectroscopy were then combined into a single
spectrum after normalizing to a common level (a correction of less then 5\%),
using a clean wavelength region 10 nm wide.  The final spectra of our stars
are shown in Figure~3.  The naming convention for our targets is
straightforward: there are three slit positions, and each position has a
pair of red and blue stars, so each name gives the color and slit position
(e.g., Blue\,2 is the blue star in the pair at slit position 2).

\begin{figure*}[t]
\begin{center}
\includegraphics[width=6.5in]{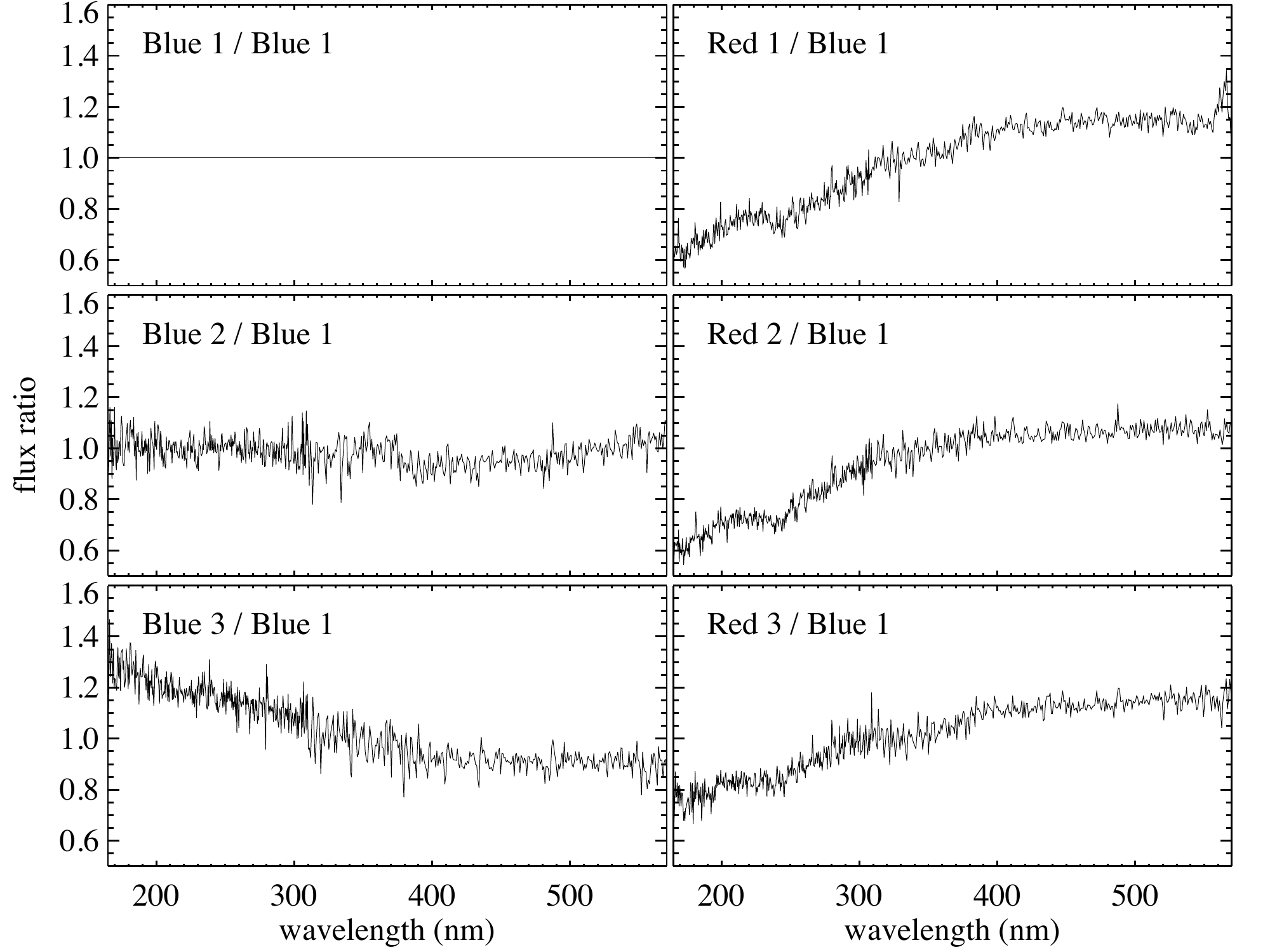}
\end{center}
\caption{The flux ratio of each spectrum to that of Blue\,1, normalized
  to unity over 310 -- 360 nm (i.e., the F336W bandpass, and just
  blueward of the Balmer jump).  Photometry of the red stars implies
  they have effective temperatures 2,000--4,000~K cooler than those of the blue
  stars.  With that temperature difference, one expects the red stars
  to have significantly redder slopes in this wavelength range (as
  evident in the {\it right panels}), but also significantly stronger
  Balmer jumps.  Instead, the flux ratio of each red star to
  Blue\,1 is surprisingly continuous across 365~nm.  A change in
  overall continuum slope without a change in Balmer jump strength
  implies that the red stars have atmospheric metallicities that are
  significantly higher than those of the blue stars (see Figure~5).  }
\end{figure*}

\begin{figure}[t]
\begin{center}
\includegraphics[width=3.2in]{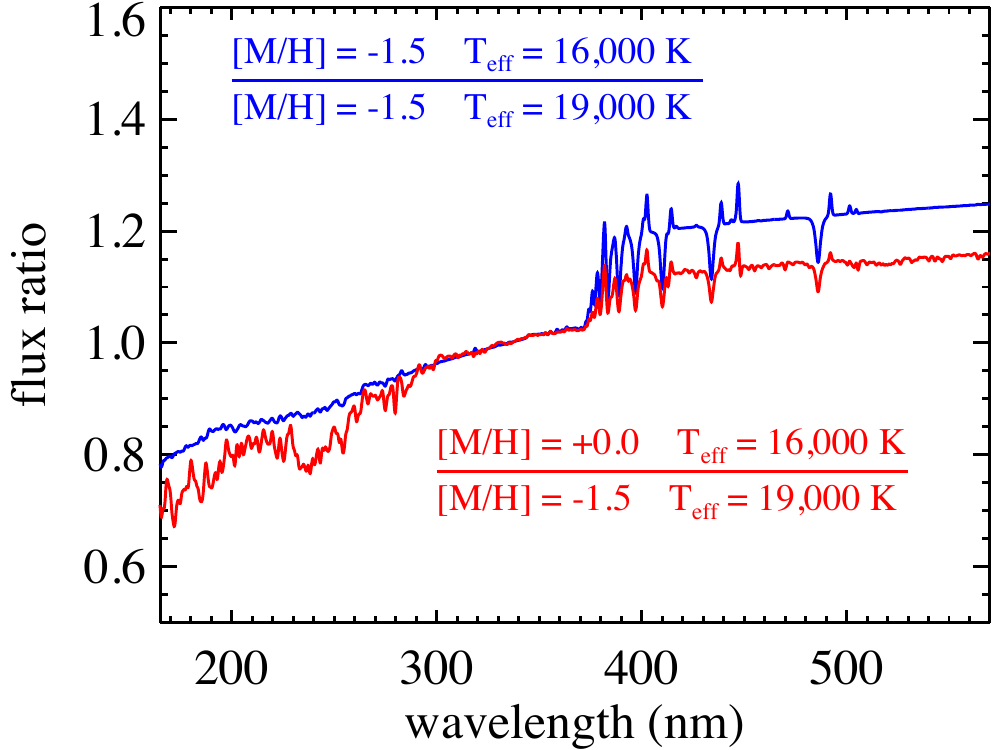}
\end{center}
\caption{The flux ratio of a 16,000~K star to a 19,000~K star ({\it blue
    curve}) exhibits a clear break at the Balmer jump when both stars
  are at the same metallicity.  When the atmospheric metallicity of the
  cooler star is elevated relative to that of the hotter star, the
  distinction in Balmer jump strength is reduced ({\it red curve}).}
\end{figure}

\section{Analysis}

To ease comparisons, we first compare each of the spectra to that of
Blue\,1 (Figure~3).  Not surprisingly, all of the red
stars have spectral slopes redder than that of Blue\,1 (or any
of the blue stars).  Photometry of our targets (Figure~1;
Bellini et al.\ 2017a) had already implied that the blue stars in our sample are
2,000--4,000~K hotter than the red stars, if one folds synthetic
spectra at the cluster metallicity through the WFC3 bandpasses.
However, the Balmer jump in these red stars is not significantly stronger
than that in the blue stars, in contradiction of expectations if the stars have the
same chemical composition.  This is clearer if we look at the ratio
of flux for each star against the flux of Blue\,1 (Figure~4).  None of
the flux ratios for the red stars show a significant break at the Balmer jump.
Assuming the red stars are truly cooler than the blue stars, the Balmer jump
in the red stars may be diminished if the red stars have higher
atmospheric metallicities than the blue stars (Figure~5). This is our first
indication of a distinction in atmospheric abundances between the groups of
stars on either side of the M-jump.

The M-jump was originally discovered (Momany et al.\ 2002) and
explored (Momany et al.\ 2004) using $U$ band photometry.  The feature
is more easily seen in the color-color diagrams of Brown et
al.\ (2016; Figure~1), which also employ a band (F336W) approximating
the $U$ band.  The phenomenon must produce spectral features that
affect the $U$ band itself or the strength of the $U$ band relative to
neighboring bandpasses.  If one considers the stars lying immediately
to the red of the M-jump in Figure~1, the photometric deviation could
be due to these stars having $m_{\rm F275W} - m_{\rm F336W}$ colors
that are too red, $m_{\rm F336W} - m_{\rm F438W}$ colors that are too
blue, or a combination thereof.  The spectra in Figure~3 imply it is
indeed due to both effects: the red stars are much redder in $m_{\rm F275W} -
m_{\rm F336W}$ color than the blue stars, but the weaker-than-expected
Balmer jumps in the red stars make their $m_{\rm F336W} - m_{\rm
  F438W}$ colors bluer than they would be otherwise.

\begin{figure*}[t]
\begin{center}
\includegraphics[width=6.5in]{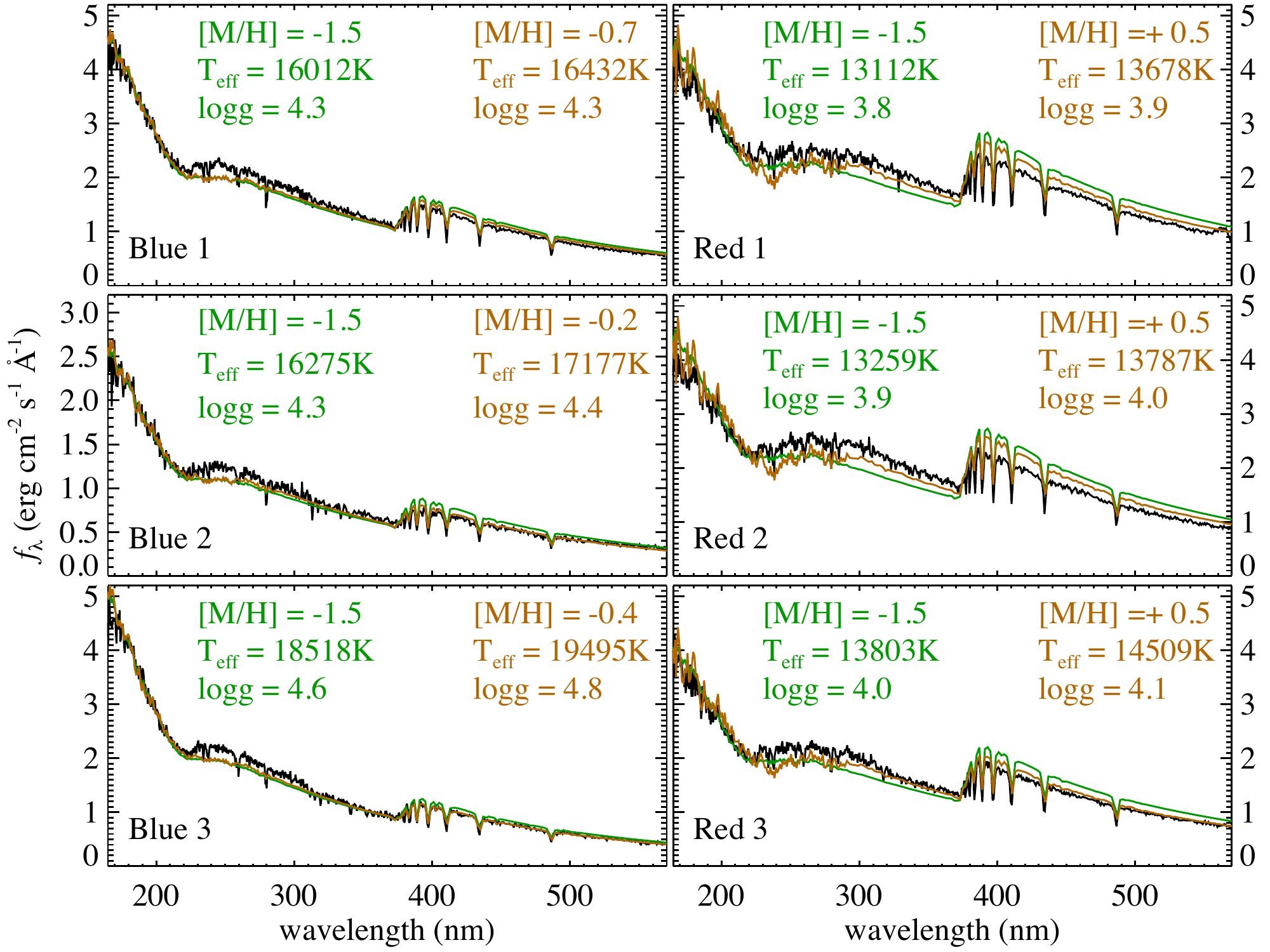}
\end{center}
\caption{The spectra of our six stars ({\it black curves}) compared to
  the best-fit synthetic spectra ({\it colored curves}) from the
  Castelli \& Kurucz (2003) grid.  In all fits, the effective
  temperature is allowed to float, and the surface gravity is fixed at
  the value corresponding to that temperature along the zero-age
  horizontal branch.  The extinction is fixed at $E(B-V) = 0.13$~mag
  (Bellini et al.\ 2017b; Milone et al.\ 2017b).
  Neither the fits with the metallicity at
  the mean cluster value ([M/H]=$-1.5$; Harris 1996; {\it green
    curves}) nor the fits with the metallicity allowed to float
  ({\it labeled; brown curves}) are able to match the general shape of
  the pseudo-continuum and the size of the Balmer jump, although the
  models with floating metallicity perform somewhat better.  Even with
  the poor fits shown here, the synthetic spectra imply that the red
  stars are $\sim$2,000~K -- 4,000~K cooler than the blue stars, and
  that the red stars have a greater enhancement of atmospheric
  metallicity due to radiative levitation.  Note that the upper metallicity
  limit in the Castelli \& Kurucz (2003) grid is $+0.5$~dex.}
\end{figure*}

To further constrain the properties of the stars straddling the
M-jump, we turn to fits of synthetic spectra to the data.  We begin
with the Castelli \& Kurucz (2003) grid of synthetic spectra.  This
grid is unlikely to satisfactorily reproduce the spectra of hot
subdwarfs, because it assumes the abundances scale together (either at
scaled solar abundance or with $\alpha$-element enhancement), while
hot subdwarfs exhibit significant atmospheric abundance anomalies
caused by atmospheric diffusion.  Nevertheless, the grid serves as a
starting point.

There are two unknowns of interest: the effective temperature and
abundance.  We cannot perform a single-parameter fit of the abundance,
because we have no independent measure of the effective temperature;
the photometry only allows an estimate of the effective temperature
given an assumed abundance.  We can perform a single-parameter fit of
the temperature, because we can assume the abundance matches that of
the cluster ([M/H]$~=-1.5$; Harris 1996).  However, as noted earlier,
the cluster hosts sub-populations with distinct metallicities (Bellini
et al.\ 2017b; Marino et al.\ 2011), while the atmospheric abundances of
hot subdwarfs show significant variations due to gravitational
settling and radiative levitation (Moehler et al.\ 1999, 2000; Behr
2003; Pace et al.\ 2006), and we do not expect these subdwarf spectra
to reflect the main sequence abundances for any of the cluster's
sub-populations.  We assume a foreground extinction of $E(B-V) =
0.13$~mag (Bellini et al.\ 2017b; Milone et al.\ 2017b).  To fit the
model to the data, we define a numerical score as the sum of the
squared differences between the observed spectrum and the model
spectrum, and then minimize this score using an amoeba simplex
algorithm.  The effective temperature is allowed to float over the
range 10,000--29,000~K, with initial estimate driven by the photometry
(under the assumption of cluster abundance);
the surface gravity is
set to that of a zero-age HB star at the given effective temperature.  The
results are shown in Figure~6 ({\it green curves}).  Although the
general shape of each spectrum can be reproduced, the region between
the 220~nm extinction bump and the 365~nm Balmer jump is
underpredicted in the models, while the size of the Balmer jump is
overpredicted in the models.  As explained above (Figure~5), for a
given effective temperature, the Balmer jump is weaker at higher
metallicity, and so it is not surprising that models at cluster
metallicity cannot match the Balmer jump in a set of subdwarfs
exhibiting significant metal enhancement from radiative levitation.
We address this concern by allowing both the temperature and
metallicity to float simultaneously
in the fit, using the same amoeba simplex algorithm to minimize
the differences between the models and data;
the results are also shown in
Figure~6 ({\it brown curves}).  Although these models are an
improvement over those at cluster metallicity, they suffer from
similar discrepancies.  That said, the models imply that the red stars
have higher atmospheric metallicities than the blue stars.  The fits
to the blue stars all favor a metallicity that is enhanced over the
cluster value, but still sub-solar.  The fits to the red stars have
metallicities consistently limited by the highest metallicity available in
the Castelli \& Kurucz (2005) grid ($+0.5$~dex).

Given the unusual atmospheric abundances of hot subdwarfs, we next
turn to models where the abundances of individual elements can be
tuned independently.  We use the {\sc atlas12} and {\sc synthe} codes of Kurucz
(2005), which allow elements to be varied independently, but we note
that these codes assume a uniform abundance of each element throughout the
atmosphere, as opposed to the stratified abundance profiles found in hot
subdwarfs (see LeBlanc et al.\ 2010).
The parameter space for such an exploration is large, and the
calculations of model atmospheres and synthetic spectra are
computationally expensive, but we can approach the problem in a manner
that limits the scope.  Michaud et al.\ (2011) calculated stellar
evolution models for HB stars at the relevant temperature.
Their models for a 14,000~K star (see their Figure~5) imply
significant enhancement for the metals Ti through Fe.  Of those
metals, Fe has the strongest spectral features in our wavelength
range, but for completeness we allow variations in Ti, V, Cr, Mn, Fe,
Co, and Ni.  Even with this limited set of abundances, it would be
prohibitively expensive to iteratively fit our spectra with a
procedure that calculated full model atmospheres and synthetic spectra
for each abundance step in the fitting process, so we break the fit
into two pieces.  First, we calculate model atmospheres and synthetic
spectra where all but one of the abundances are at the cluster
metallicity, and one of these seven abundances is stepped over the
cluster value by an enhancement of 0 to $+4.5$~dex with 0.5~dex steps
(i.e., abundances
from $-1.5$ to $+3.0$ dex).  This provides an approximation for the
spectral features of each individual element as a function of abundance.
We then construct a
model spectrum from a baseline spectrum (all elements at cluster
abundance) and a linear combination of the features from each of our
varied elements, determined by comparison to our data with 
minimization of differences.
Although this linear combination of spectral features is
a poor approximation for the true behavior in a stellar atmosphere,
the approximation is sufficient for a first estimate.  The resulting
abundances are then used to construct model atmospheres and associated
synthetic spectra, with these spectra serving as the new baseline in
the next iteration of fits.  The baseline is again modified by a
linear combination of spectral features from our seven varied elements,
with the results used to calculate the next set of model atmospheres
and synthetic spectra.  The iterations are continued until the
abundance changes for each element are less than 0.1~dex between
iterations, at which point the model is considered converged.  By the
end of the iterations, the spectral modifications over the baseline,
approximated from the library of spectral features for each element, are
a small perturbation.  After convergence, a final model atmosphere and
synthetic spectrum is calculated for each star (Figure~7); the abundance
values are plotted in Figure~8 to ease comparisons.
The agreement between synthetic and observed spectrum is much improved
over the fits with the Castelli \& Kurucz (2003) grid, but there are
still discrepancies: a residual mismatch in the Balmer jump and
deviations at $\sim$210--260~nm (a region associated with strong Fe
features).  The effective temperatures implied by our spectroscopic
fits are within 600~K of the temperatures implied by the photometry
when that photometry is aligned to zero-age HB models (Brown et al.\ 2016).

\begin{figure*}[t]
\begin{center}
\includegraphics[width=6.5in]{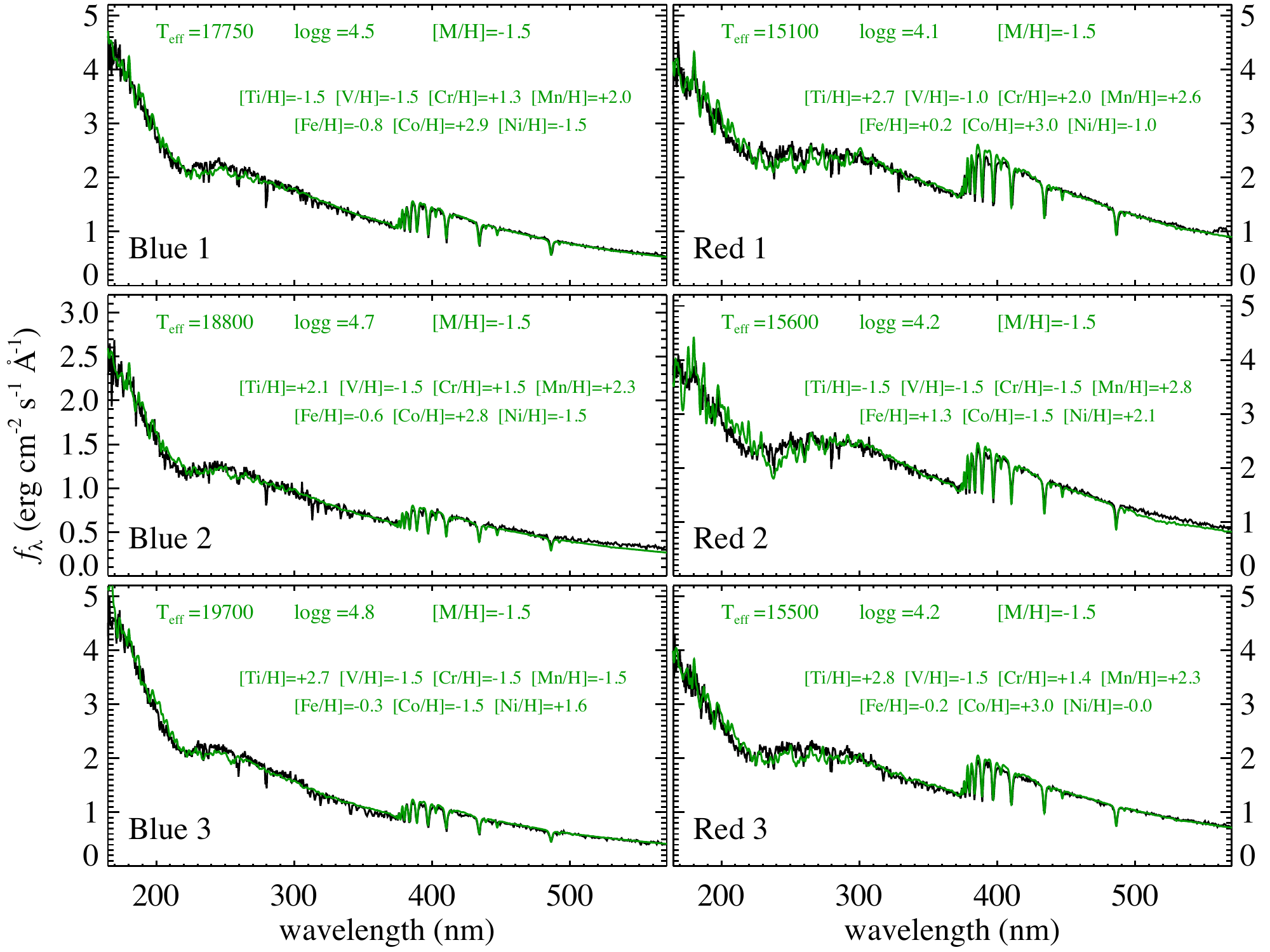}
\end{center}
\caption{The spectra of our six stars ({\it black curves}) compared to
  the best-fit synthetic spectra ({\it green curves}) from {\sc atlas12} and
  {\sc synthe}.  In these models, most elements are held at the mean cluster
  abundance ([M/H]=$-1.5$; Harris 1996), but 7 elements ({\it labels})
  near the Fe peak are allowed to vary up to an enhancement of
  4.5~dex over the cluster value, because they can be significantly
  affected by radiative levitation (Michaud et al.\ 2011).
  The best-fit values in each fit ({\it labels}) are
  shown in Figure~8 to ease comparisons.
  The changes induced on each spectrum by the enhanced atmospheric
  metallicities are much larger in the red stars than the blue stars
  (Figure~9).}
\end{figure*}

\begin{figure}[t]
\begin{center}
\includegraphics[width=3.2in]{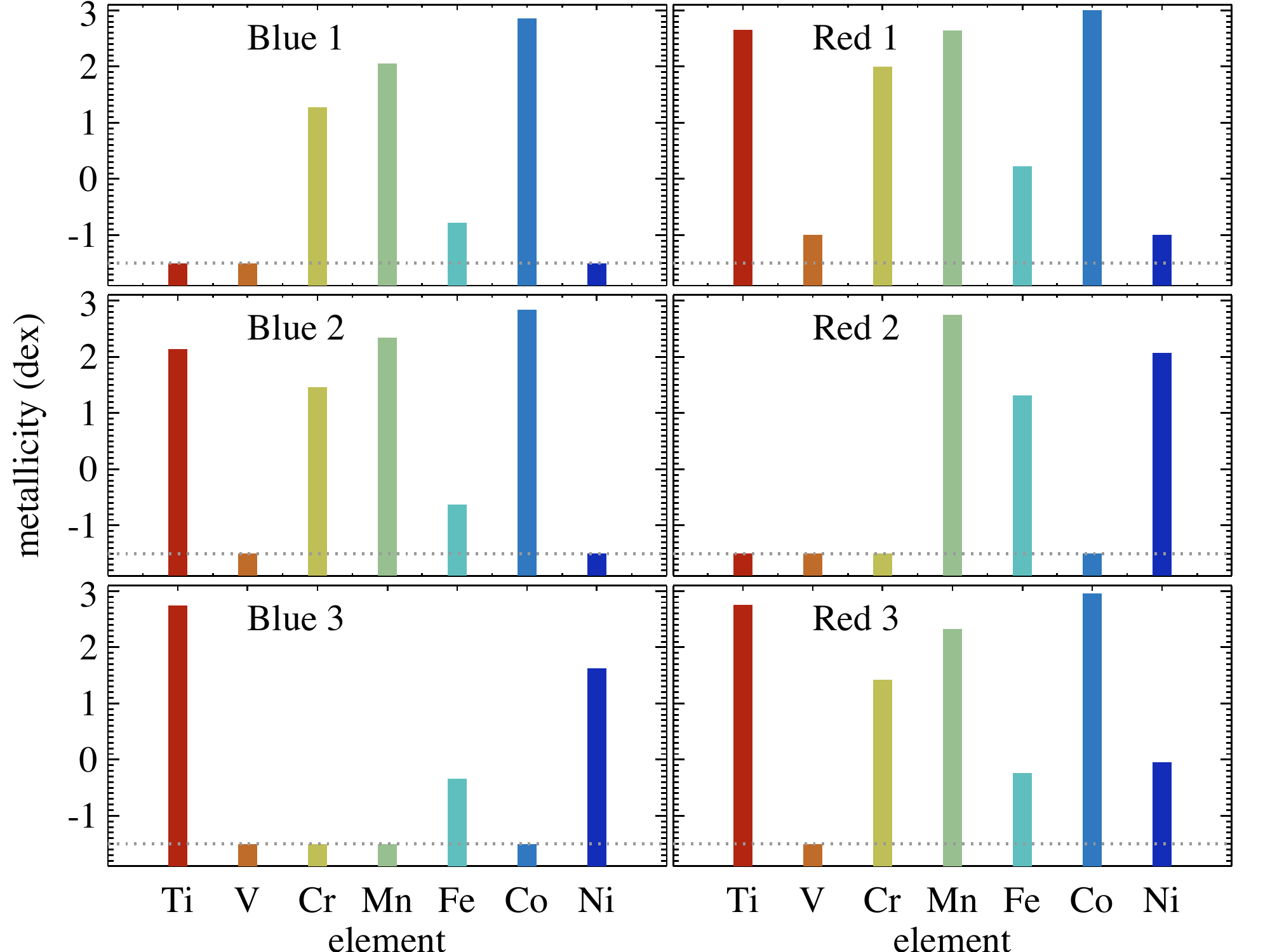}
\end{center}
\caption{The abundance enhancements ({\it colored bars}) in the
  best-fit model for each star (Figure~7).  All of the stars exhibit
  Fe abundances elevated over the cluster value ({\it dotted
    line}), but the red
  stars all have higher Fe abundances than the blue stars (albeit
  with significant star-to-star variation).  Ni, Mn,
  and Cr have a second-order effect on the spectra; these elements
  tend to exhibit higher enhancements in the red stars than the blue
  stars.
}
\end{figure}

For several reasons, it is difficult to provide formal uncertainties
for our abundances.  Due to the heavy line-blanketing in this
wavelength range and the low-resolution gratings employed in the
observations, none of the elemental abundances in our fits are
associated with a well-defined absorption line (or set of lines) in
our spectra. Instead, there are regions of the spectra affected to
varying degrees by the abundance of a particular element, with no
clear region that should be defined to evaluate the fit for each
element.  Given the
overlapping features of these elements, variations in one elemental
abundance can be partly compensated by a combination of changes in
other elements. Moreover, the processing of the individual raw counts exposures
into the final spectrum for each star involves various data reduction
packages that do not propagate the uncertainties in each spectral bin;
furthermore, given the finite line spread function and the spatial
dithering, the uncertainties in the spectral bins are correlated on
the scale of a few bins.  Finally, the human and computer labor
associated with the production of model atmospheres and synthetic
spectra makes it highly impractical to explore the confidence contours
in the 7-dimensional parameter space of the abundances varied.
However, we can crudely estimate the uncertainties in our abundances
by varying individual abundances to trial levels, allowing the other
elements to refit and compensate, and inspecting to see if the result is
a noticeably worse match to the data.  Varying the abundances of Fe,
Cr, Mn, and Co implies they are uncertain at the level $\sim$0.2~dex,
while the abundances of Ti and Ni are uncertain at the level of
$\sim$0.5~dex.  The uncertainty in the V abundance is larger than
1~dex (unsurprising, given its weak features).  Despite these
uncertainties, our analysis indicates that the M-jump is due
to a lower atmospheric Fe abundance in the stars hotter than the jump.

\begin{figure*}[t]
\begin{center}
\includegraphics[width=6.5in]{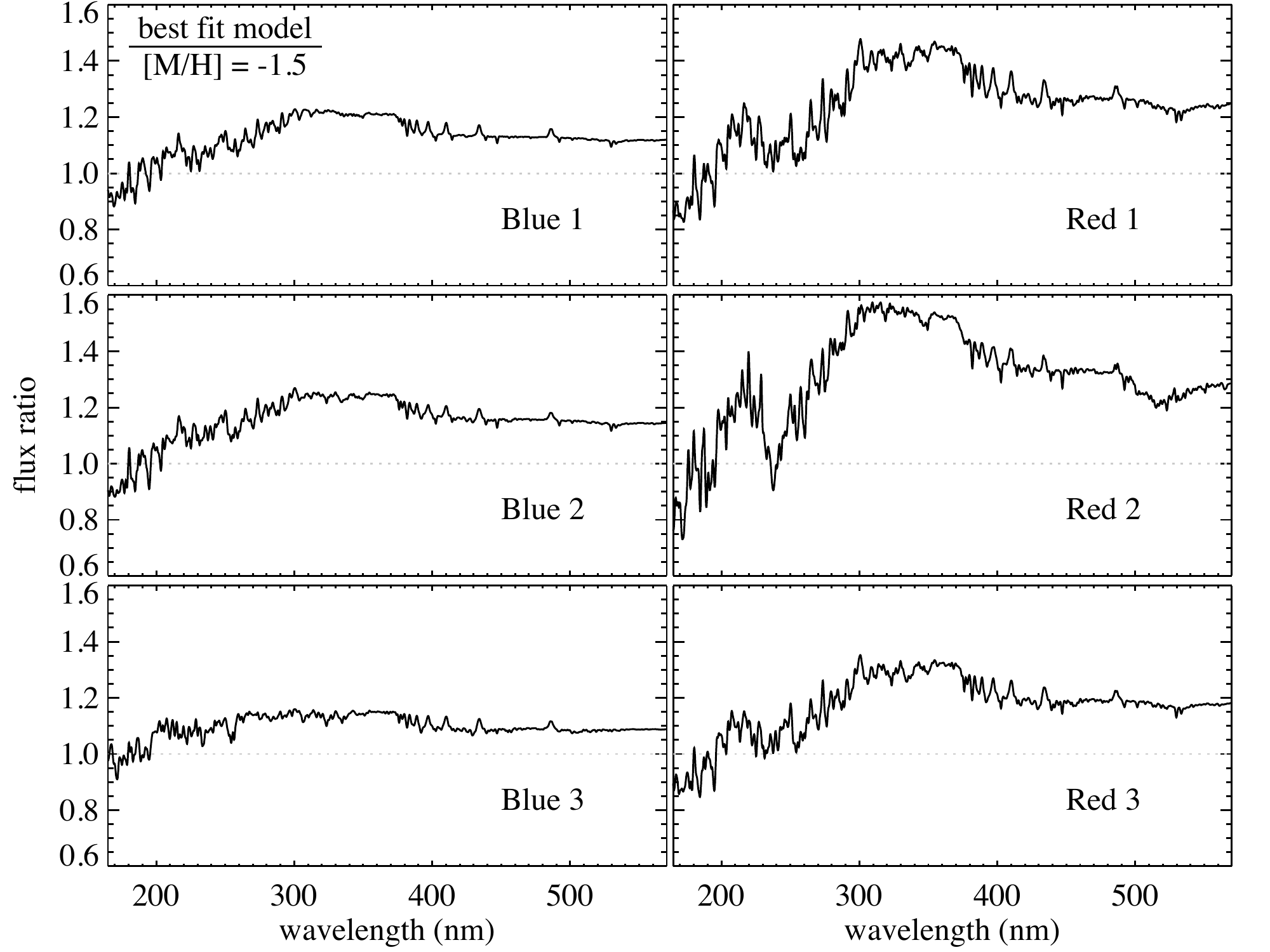}
\end{center}
\caption{The flux ratio of each best-fit synthetic spectrum
  (Figure~7) to a synthetic spectrum at the same temperature and gravity, but
  mean cluster metallicity ([M/H]$=-1.5$). The spectral features in the
  red stars are all significantly stronger than those in the blue stars,
  primarily due to the higher Fe abundances and lower effective temperatures
  in the red stars.  
}
\end{figure*}
  
\begin{figure*}[t]
\begin{center}
\includegraphics[width=6.5in]{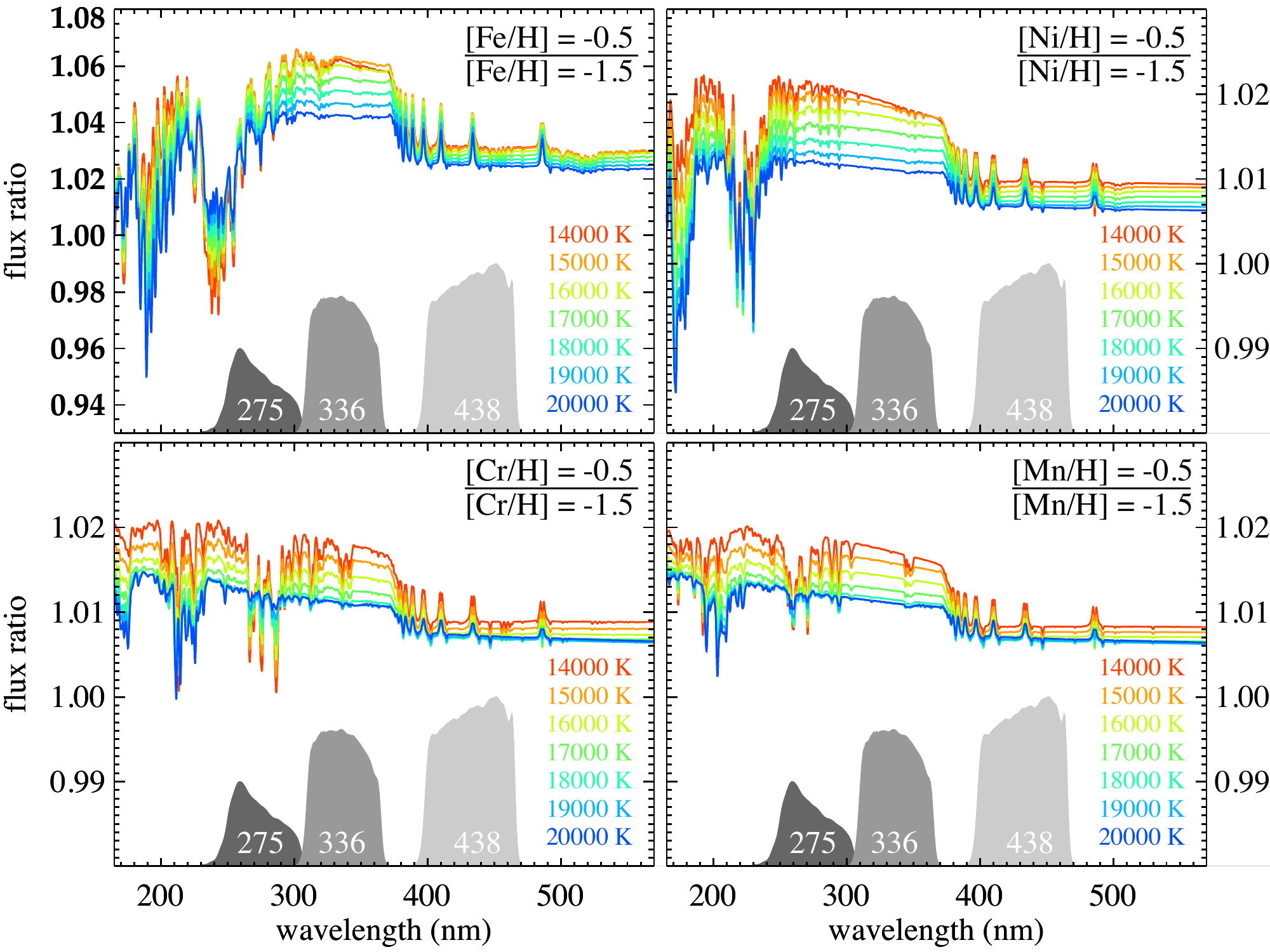}
\end{center}
\caption{The effect on the spectrum of a hot subdwarf when one of 4
  abundances ({\it labeled}) is enhanced by 1~dex, as a function of
  effective temperature.  Of the elements varied in our fits, Fe ({\it
    upper left}) has the largest impact in the wavelength range spanned
  by the F275W, F336W, and F438W filters (bandpasses shown here for
  reference at arbitrary normalization, and vertically
  shifted to the bottom of the plot; {\it grey shading}),
  while Cr, Mn, and Ni have a second-order effect (note that the
  ordinate scale for these three elements is three times smaller than
  that for Fe).  Each curve shows the flux ratio for the synthetic
  spectrum at enhanced metallicity to the spectrum at mean cluster
  metallicity, with the colors indicating effective temperature ({\it
    labeled}).  The effect of enhanced metallicity is more significant
  at the temperatures of the red stars in our sample.  For Fe ({\it
    upper left}), the flux falling in the F336W bandpass, relative to
  that falling in the F275W and F438W bandpasses, is a particularly
  strong function of effective temperature.  }
\end{figure*}

There are three clear results in these fits: the blue stars are
$\sim$2,000--4,000~K hotter than the red stars, all of the stars
exhibit abundance enhancements over the cluster value, and the Fe
enhancement in the red stars is larger than that in the blue stars.
The larger Fe enhancements in the red stars agree with the 
sense of the M-jump in the color-color
plane (Figure~1); our synthetic spectra imply that an increase in Fe
abundance by 1~dex will shift the $m_{\rm F275W} - m_{\rm F438W}$ color
0.01~mag redder and increase the $C$ color index by 0.07~mag.
The other abundances in our fit do not provide a wholly consistent
story (Figure~8), but among those that have a secondary influence (Ni, Mn, and
Cr), there is a weak trend (with exceptions) for the red stars
to have higher enhancements than the blue stars.
The remaining elements (Ti, V, and Co) have an even
smaller effect in this wavelength range, and there is no clear pattern
for these elements.  Given this behavior, we also explored
models where the abundances of all seven elements were varied in unison;
the resulting models agreed less well with the data (as one would expect,
given fewer free parameters and the unphysical congruence of abundances),
but still implied larger abundance enhancements in the red stars.

Relative to spectra with cluster abundances, the red stars clearly
exhibit larger spectral deviations than the blue stars.  This can be seen in
Figure~9, which shows the ratio of flux in the best-fit synthetic
spectrum to that from a model with cluster abundances at the same
effective temperature and surface gravity.  The larger deviations in
the red spectra are driven by the larger Fe abundances
(Figure~8) and by the fact that the metals produce stronger
spectral features at the lower effective temperatures of the red
stars (Figure~10).

\section{Discussion}

The HB distributions of globular clusters exhibit discontinuities at
$\sim$12,000~K and $\sim$18,000~K that are remarkably consistent in
effective temperature.  The cooler G-jump remains constant despite
large changes in many population parameters, with the exception of
populations greatly enhanced in helium; the hotter M-jump has
exhibited no variation in any cluster observed to date, regardless of
population parameters (Brown et al.\ 2016).  Although the G-jump
has been attributed to the onset of radiative levitation in hot
subdwarfs, the origin of the M-jump has remained unclear.  Both
features reflect universal phenomena in the atmospheres of hot
subdwarfs, and are thus useful fiducials when comparing
photometric datasets from distinct populations.

Our spectra of stars straddling the M-jump in $\omega$~Cen imply that
the feature is due to a change in the atmospheric Fe abundance.  While
stars on either side of the M-jump exhibit greatly enhanced metal
abundances, the enhancement of Fe is stronger for stars falling to the
red side of the M-jump, relative to those stars on the blue side of
the M-jump.  The effect is amplified by the fact that the Fe spectral
features are a function of effective temperature, and stronger for the
cooler stars on the red side of the M-jump.  The high Fe abundance
produces strong spectral features within the F275W bandpass, making
the $m_{\rm F275W} - m_{\rm F336W}$ color redder, and weakens the Balmer jump,
making $m_{\rm F336W} - m_{\rm F438W}$ color bluer, producing the deviation
observed in the color-color diagram of Figure~1.

The same phenomenon appears to be at work in the hot subdwarfs of the
Galactic field.  Geier et al.\ (2010; their Figure~1) show the Fe
abundance for field subdwarfs spanning the effective temperatures of
both the G-jump and the M-jump.  The clear rise in Fe abundance at the
G-jump ($\sim$12,000~K) is the most striking feature in their figure,
but there appears to be a drop in Fe abundance when one compares stars
on either side of M-jump ($\sim$18,000~K); the drop is not
significant enough to warrant discussion in their paper, and the dearth
of stars near 18,000~K makes it difficult to see if the transition
is sharp.  

The change in Fe abundance at the M-jump appears to be supported by
radiative levitation modeling, as well. Michaud et al.\ (2011) calculated
radiative accelerations and associated abundance profiles for metals
in the atmospheres of hot subdwarfs spanning 10,700--30,400~K.  The Fe surface
abundances in their 10,700 and 15,000~K models are significantly higher than that
in their 20,000~K model, which is in turn higher than those in their 25,000 and
30,400~K models (see their Figure~8).  Like our spectra, their models
imply that Fe remains enhanced at temperatures hotter than the M-jump, but
that the enhancements are not as strong as those at cooler temperatures.
In the Michaud et al.\ (2011) calculations,
the surface abundance enhancements in the various elements do not conform
to a regular pattern, given the interplay between depth-dependent
radiative acceleration, temperature, and line saturation; for this reason
the physical mechanism for the break in the Fe enhancement, responsible for the
M-jump, is unclear.  The models do not give much insight into the large
star-to-star Fe variations in our sample. The radiative levitation timescales
are a small fraction of the HB lifetime, such that most HB stars should
be observed after the enhancements have stabilized.  Brown et al.\ (2016)
also noted that the \ion{He}{2} convection zone encroaches upon the surface
near the effective temperature of the M-jump, and this encroachment may
play a role in the abundance changes producing the M-jump, similar to the
role played by convection in the G-jump.

\acknowledgements

Support for program GO-14759 was provided by NASA through a grant from
the Space Telescope Science Institute, which is operated by the
Association of Universities for Research in Astronomy, Inc., under
NASA contract NAS 5-26555.  We are grateful to F.\ Castelli for making
available the model atmosphere ({\sc atlas12}) and synthetic spectra
({\sc synthe}) codes of R.\ Kurucz in formats compatible with modern
compilers, and for her assistance with the installation of these
codes.  We also thank R.\ Kurucz for making his codes publicly
available.  S.\ Lockwood kindly answered STIS calibration questions
and fixed a bug in the {\sc stis\_cti} software.  We appreciate useful
discussions with A.\ Aparicio.

\end{document}